\documentclass{article}
\usepackage{amsmath, amssymb, amsfonts}
\title{Negative $\Lambda$ induced by accelerated motion}  
\author{Hristu Culetu, \\Ovidius University, Dept.of Physics, \\ Mamaia Avenue 124, 900527 Constanta, Romania, \\e-mail : hculetu@yahoo.com}

\begin{document}
\numberwithin{equation}{section}
\pagenumbering{arabic}
\maketitle
\newcommand{\fv}{\boldsymbol{f}}
\newcommand{\tv}{\boldsymbol{t}}
\newcommand{\gv}{\boldsymbol{g}}
\newcommand{\OV}{\boldsymbol{O}}
\newcommand{\wv}{\boldsymbol{w}}
\newcommand{\WV}{\boldsymbol{W}}
\newcommand{\NV}{\boldsymbol{N}}
\newcommand{\hv}{\boldsymbol{h}}
\newcommand{\yv}{\boldsymbol{y}}
\newcommand{\RE}{\textrm{Re}}
\newcommand{\IM}{\textrm{Im}}
\newcommand{\rot}{\textrm{rot}}
\newcommand{\dv}{\boldsymbol{d}}
\newcommand{\grad}{\textrm{grad}}
\newcommand{\Tr}{\textrm{Tr}}
\newcommand{\ua}{\uparrow}
\newcommand{\da}{\downarrow}
\newcommand{\ct}{\textrm{const}}
\newcommand{\xv}{\boldsymbol{x}}
\newcommand{\mv}{\boldsymbol{m}}
\newcommand{\rv}{\boldsymbol{r}}
\newcommand{\kv}{\boldsymbol{k}}
\newcommand{\VE}{\boldsymbol{V}}
\newcommand{\sv}{\boldsymbol{s}}
\newcommand{\RV}{\boldsymbol{R}}
\newcommand{\pv}{\boldsymbol{p}}
\newcommand{\PV}{\boldsymbol{P}}
\newcommand{\EV}{\boldsymbol{E}}
\newcommand{\DV}{\boldsymbol{D}}
\newcommand{\BV}{\boldsymbol{B}}
\newcommand{\HV}{\boldsymbol{H}}
\newcommand{\MV}{\boldsymbol{M}}
\newcommand{\be}{\begin{equation}}
\newcommand{\ee}{\end{equation}}
\newcommand{\ba}{\begin{eqnarray}}
\newcommand{\ea}{\end{eqnarray}}
\newcommand{\bq}{\begin{eqnarray*}}
\newcommand{\eq}{\end{eqnarray*}}
\newcommand{\pa}{\partial}
\newcommand{\f}{\frac}
\newcommand{\FV}{\boldsymbol{F}}
\newcommand{\ve}{\boldsymbol{v}}
\newcommand{\AV}{\boldsymbol{A}}
\newcommand{\jv}{\boldsymbol{j}}
\newcommand{\LV}{\boldsymbol{L}}
\newcommand{\SV}{\boldsymbol{S}}
\newcommand{\av}{\boldsymbol{a}}
\newcommand{\qv}{\boldsymbol{q}}
\newcommand{\QV}{\boldsymbol{Q}}
\newcommand{\ev}{\boldsymbol{e}}
\newcommand{\uv}{\boldsymbol{u}}
\newcommand{\KV}{\boldsymbol{K}}
\newcommand{\ro}{\boldsymbol{\rho}}
\newcommand{\si}{\boldsymbol{\sigma}}
\newcommand{\thv}{\boldsymbol{\theta}}
\newcommand{\bv}{\boldsymbol{b}}
\newcommand{\JV}{\boldsymbol{J}}
\newcommand{\nv}{\boldsymbol{n}}
\newcommand{\lv}{\boldsymbol{l}}
\newcommand{\om}{\boldsymbol{\omega}}
\newcommand{\Om}{\boldsymbol{\Omega}}
\newcommand{\Piv}{\boldsymbol{\Pi}}
\newcommand{\UV}{\boldsymbol{U}}
\newcommand{\iv}{\boldsymbol{i}}
\newcommand{\nuv}{\boldsymbol{\nu}}
\newcommand{\muv}{\boldsymbol{\mu}}
\newcommand{\lm}{\boldsymbol{\lambda}}
\newcommand{\Lm}{\boldsymbol{\Lambda}}
\newcommand{\opsi}{\overline{\psi}}
\renewcommand{\tan}{\textrm{tg}}
\renewcommand{\cot}{\textrm{ctg}}
\renewcommand{\sinh}{\textrm{sh}}
\renewcommand{\cosh}{\textrm{ch}}
\renewcommand{\tanh}{\textrm{th}}
\renewcommand{\coth}{\textrm{cth}}

\begin{abstract}
A correlation between accelerated motion and a noncompact 5th dimension is proposed. The curvature invariants and the stress energy tensor in the bulk depend only on the 5th dimension $w$ and vanish asymptotically while the proper acceleration of a static observer is proportional to $1/w$. The brane (located at $w = w_{0}$) metric is conformally flat (of AdS type) with $\Lambda = -3/w_{0}^{2}$ and the invariant acceleration $a = \sqrt{a^{b}a_{b}} = 1/w_{0}$. Therefore, we assume that a hyperbolic observer with the rest-system acceleration $a$ is embedded in the 5th dimension at $w_{0} = 1/a$.
 \end{abstract}
 
 \section{Introduction}
 Despite the great successes of the Standard Model in particle physics and cosmology, there are still serious problems which have to be solved, such as the cosmological constant (c.c) problem, the black hole information loss paradox, the dark energy problem and the fact that our universe is accelerating. As Stojkovic \cite{DS} has noticed, some radical new ideas are needed to face those problems. Instead of changing the known physical theories, we could use a different background to formulate them. Stojkovic proposed that the space is lower dimensional at scales shorter than $10^{-17} cm$ and is higher dimensional at scales larger than one Gpc \cite{DS} (see also \cite{DS1, LA}). In other words, the number of dimensions increases with the scale. If so, there is no need to quantize the $3+1$-dimensional gravity but instead we should quantize $2+1$-dimensional gravity.
 
 Changing the dimensions of the universe at large distances may have consequences for cosmology. Ponce de Leon \cite{JPL} and Overduin and Wesson \cite{OW} focused on an homogeneous and isotropic solution of vacuum Einstein's equations in five dimensions
      \begin{equation}
  ds^{2} =- dt^{2} + e^{2\sqrt{\Lambda/3}t} (dr^{2}  + r^{2} d \Omega^{2}) + dw^{2},
 \label{1.1}
 \end{equation}
where $\Lambda = 3/w^{2}$ and $d \Omega^{2}$ stands for the metric on the unit 2-sphere. On the $w = const.$ hypersurfaces, the above geometry reduces to a 4-dimensional de Sitter geometry with constant $\Lambda$ and, therefore, an observer located on a $w = const.$ slice will measure a nonzero stress tensor. This ''induced matter'' energy-momentum tensor arises from pure geometry in the higher dimensional space and is sometimes called ''shadow matter'' \cite{FFS}. We see that observers located at different slices of 5-dimensional space acquire different values of the c.c. Izumi and Shiromizu \cite{IS} remarked that a spacetime with compact extra dimensions is semiclassically unstable. The space decays to the Witten bubble-type space (''bubble of nothing'') \cite{EW, BM, CJ, HC1}. They constructed exact solutions with the Witten bubble in the DGP \cite{DGP} braneworld model  and found that the geometry on the single brane looks like an Einstein-Rosen bridge. 

We develop in this work a solution of 5-dimensional Einstein's equations with a stress tensor satisfying the equation of state $p = -\rho$ on the hypersurface $w = w_{0} = const.$, where $w$ is the 5th dimension. The metric coefficients on the 4-dimensional subspace depend on $w$ which is viewed as an inverse acceleration. The 5-dimensional space is conformally-flat and all curvature invariants depend on $w$ only, being divergent at $w = 0$. Applying the Shiromizu et al. prescription \cite{SMS} (see also \cite{MS}) upon the Gauss-Codazzi equations for the variation of the extrinsic curvature, we have written down the stress energy tensor on the brane $w = w_{0}$  and found that the 5-dimensional Newton's constant depends on which hypersurface we are located in. In addition, we proposed that an accelerated observer finds himself in a curved (AdS) spacetime with a negative cosmological constant generated by acceleration.

We use everywhere the velocity of light $c = 1$, unless explicitly stated.

\section{Bulk geometry}
In our framework, the 3-dimensional sheet we are located is embedded in a ($4+1$)- dimensional bulk, whose geometry we propose to be of the form
 \begin{equation}
  ds^{2} = \frac{w^{2}}{z^{2}}(- dt^{2} + dx^{2} + dy^{2} + dz^{2}) + dw^{2},
 \label{2.1}
 \end{equation}
 where $t, x, y, z$ are the common Cartesian coordinates and $w$ stands for the fifth dimension which will be considered to be noncompact. We take from now on $z \geq 0,~w\geq 0$, for simplicity.
 
 The $w = const.$ hypersurfaces represent an anti de Sitter (AdS) spacetime and has been studied elsewhere \cite{HC2}, but in spherical coordinates. To be an exact solution of Einstein's equation in five dimensions, $^{5}G_{ab} = 8\pi G_{5}~^{5}T_{ab}$, we need the following source on its r.h.s.
 \begin{equation}
 8\pi G_{5}~^{5}T^{a}_{~b} = diag\left(\frac{6}{w^{2}}, \frac{6}{w^{2}}, \frac{6}{w^{2}}, \frac{6}{w^{2}}, \frac{12}{w^{2}}\right). 
 \label{2.2}
 \end{equation} 
 where $G_{5}$ is the 5-dimensional Newton constant. We notice that $^{5}T_{ab}$ corresponds to an anisotropic fluid with
  \begin{equation}
  -\rho = p_{x} = p_{y} = p_{z} = \frac{p_{w}}{2} = \frac{6}{w^{2}}
 \label{2.3}
 \end{equation} 
 where $\rho < 0$ is the fluid energy density and $p_{x}, p_{y}, p_{z}, p_{w}$ are the corresponding pressures. These kinematical parameters are divergent at $w = 0$ and vanish asymptotically. Calculating the curvature invariants of the metric (2.1) we find that the scalar curvature, the Kretschmann invariant and the Ricci tensor squared are, respectively, given by
   \begin{equation}
    ^{5}R^{a}_{~a} = -\frac{24}{w^{2}},~~~ ^{5}R^{abcd}~ ^{5}R_{abcd} = \frac{96}{w^{4}},~~~ ^{5}R^{ab}~ ^{5}R_{ab} = \frac{144}{w^{4}}.
 \label{2.4}
 \end{equation} 
 All the above invariants are divergent on the hypersurface $w = 0$ which from (2.1) seems to represents a null surface. On the other hand, the invariants vanish asymptotically (same for the components of $^{5}T^{a}_{~b}$) which leads us to the conclusion that $w \rightarrow \infty$ corresponds to the 4-dimensional Minkowski spacetime. In addition, the Weyl tensor $C^{a}_{~bcd} = 0$ so the geometry (2.1) is conformally-flat. 
 
 Let us take now a congruence of static observers in the geometry (2.1), with the velocity vector field
    \begin{equation}
    u^{a} = (\frac{z}{w}, 0, 0, 0, 0)
 \label{2.5}
 \end{equation} 
with $u^{b} u_{b} = -1$. The acceleration 4-vector $a^{b} = u^{a} \nabla_{a} u^{b}$ has the nonzero components
    \begin{equation}
    a^{z} = - \frac{z}{w^{2}},~~~a^{w} = \frac{1}{w},
 \label{2.6}
 \end{equation} 
 with the proper acceleration
 \begin{equation}
 A \equiv \sqrt{a^{b}a_{b}} = \frac{\sqrt{2}}{w}.     
 \label{2.7}
 \end{equation} 
 The acceleration is vanishing when $w \rightarrow \infty$, i.e. the static observer is geodesic, as it should be in the Minkowski spacetime.
 
 \section{Brane-world stress tensor}
 We take for the time being a general bulk spacetime with five dimensions. Our 4-dimensional world is described by a three-brane embedded in a 5-dimensional space. Let $n^{a}$ be the spacelike unit vector field normal to the brane hypersurface and $h_{ab} = g_{ab} - n_{a}n_{b}$, the induced metric on the brane ($g_{ab}$ is the full 5-dimensional metric). Shiromizu et al. \cite{SMS} have shown that, from the Gauss equations relating the Riemann tensors in 5- and 4-dimensions and the Codazzi equations for the variation of the extrinsic curvature, one readily obtains
  \begin{equation}
  \begin{split}
  G_{ab} = (^{5}R_{cd} - \frac{1}{2}g_{cd}~ ^{5}R)h^{c}_{~a}h^{d}_{~b} + ^{5}R_{cd}n^{c}n^{d}h_{ab} + K_{ab} K - K^{c}_{a}K_{bc}\\ - \frac{1}{2}h_{ab}(K^{2} - K^{cd}K_{cd}) - E_{ab}
  \end{split}
 \label{3.1}
 \end{equation} 
where $K = K^{a}_{~a}$ is the trace of the extrinsic curvature $K_{ab} = h^{c}_{a}h^{d}_{b}\nabla_{c}n_{d}$, $G_{ab}$ is the 4-dimensional Einstein tensor and 
 \begin{equation}
 E_{ab} = ^{5}C_{acbd}n^{c}n^{d}.
 \label{3.2}
 \end{equation} 
We wish now to apply the Shiromizu et al. formalism \cite{SMS} (see also \cite{MS}) for the 5-dimensional metric (2.1). We choose, for convenience, the brane to be located on the hypersurface $w = w_{0}$, so that the normal to the brane is $n^{a} = (0, 0, 0, 0, 1)$. We now evaluate the terms from the r.h.s. of (3.1). Noting firstly that $E_{ab} = 0$ since (2.1) is conformally-flat. For the nonzero components of the extrinsic curvature of the $w = w_{0}$ hypersurface one obtains
 \begin{equation}
 K_{xx} =  K_{yy} =  K_{zz} = - K_{tt} = \frac{w_{0}}{z^{2}}
 \label{3.3}
 \end{equation} 
with $K^{a}_{a} = \nabla_{a}n^{a} = 4/w_{0}$. Keeping in mind that 
 \begin{equation}
 ^{5}R^{a}_{~b} = diag\left(-\frac{6}{w^{2}}, -\frac{6}{w^{2}}, -\frac{6}{w^{2}}, -\frac{6}{w^{2}}, 0\right), 
 \label{3.4}
 \end{equation} 
we have on the brane
 \begin{equation}
 G^{a}_{~b} = diag\left(\frac{3}{w_{0}^{2}}, \frac{3}{w_{0}^{2}}, \frac{3}{w_{0}^{2}}, \frac{3}{w_{0}^{2}}\right). 
 \label{3.5}
 \end{equation} 
We use now the Lanczos equation
  \begin{equation}
  -K_{ab} + h_{ab}K = LG_{ab} = 8\pi G_{5}T_{ab},
 \label{3.6}
 \end{equation} 
where $L$ has a length scale and $T_{ab}$ is the stress tensor on the brane. To find how $G_{5}$ and $G_{4}$ are correlated in Eq. (3.6), we employ (3.3) and (3.5) in (3.6) and obtain $L = w_{0}$. We get now $G_{4} = G_{5}/w_{0}$. 

Keeping in mind (2.7), it is suitable to read $1/w_{0}$ as a constant acceleration, namely, $w_{0} = 1/a$. With this choice, the metric on the brane appears as
  \begin{equation}
  ds^{2} = \frac{1}{a^{2}z^{2}}(- dt^{2} + dx^{2} + dy^{2} + dz^{2}),
 \label{3.7}
 \end{equation}
which is an AdS spacetime written in Cartesian coordinates, with $\Lambda = -3a^{2}$. Eq. (3.7) is just the metric (2.2) from \cite{HC2} when the mass $m = 0$ and the transformation from the Cartesian coordinates to spherical coordinates is performed. 

The curvature invariants on the brane are given by
   \begin{equation}
   R^{a}_{~a} = -12a^{2},~~~ R^{abcd}R_{abcd} = 24a^{4},~~~ R^{ab}R_{ab} = 36a^{4},
 \label{3.8}
 \end{equation} 
with a vanishing Weyl tensor. We note also that all the curvature invariants do not depend on the coordinate $z$ (same is valid for the stress tensor). The velocity vector field $u^{a} = (az, 0, 0, 0)$ of a static observer leads to the acceleration
   \begin{equation}
   a^{b} = (0, 0, 0, -a^{2}z),
 \label{3.9}
 \end{equation} 
with $\sqrt{a^{b}a_{b}} = a$. In other words, our interpretation of $w_{0}$ as an inverse acceleration seems to be correct. We read this in the following manner: when an observer undergoes a hyperbolic motion with constant acceleration $a$, he/she immerses in the fifth dimension at $w = w_{0} = 1/a$. When $a$ varies, so does the observer's position in the fifth dimension. Our approach is similar with that of Billyard and Sajko \cite{BS} to whom the 5-dimensional spacetime can be viewed as a foliation of the 4-dimensional sheet or of D.-C. Dai et al. \cite{DSWZ} (see also \cite{HS}) who consider that our (3+1)-dimensional universe is actually a dense stack of multiple parallel (2+1)-dimensional branes (they generalize the DGP model to a multibrane case). 

We must remind that $a$ is the rest-system acceleration, i.e. observer's acceleration w.r.t. an inertial observer instantaneously at rest with the accelerating one. The difference is that for the inertial observer $w_{0} = \infty$ but $w_{0} = 1/a$ for the accelerating observer. Hence, even though at some instant they have the same 4-dimensional coordinates and velocities, they are immersed at different values of $w$.

Let us recall Overduin's and Wesson's conjecture that $w$ could be taken as the particle mass, namely $w = Gm/c^{2}$. That allows us to treat the rest mass $m$ as a length coordinate in analogy with $x^{0} = ct$ \cite{OW}. A comparison with our choice gives $a = c^{4}/Gm$, namely $a \propto \kappa$, where $\kappa = c^{4}/4Gm$ is the surface gravity associated to the mass $m$ (if it were a black hole). However, our proposal is not related to any mass. As we have seen in a previous paper \cite{HC2}, inertial forces arise from the curvature induced by the negative c.c. $\Lambda = -3a^{2}$.

\section{Conclusions}
It is clear that some radical ideas are necessary to face fundamental problems as the c.c. problem, the accelerating universe, the black hole information issue or the quantum gravity conundrum. Ponce de Leon \cite{JPL} and Overduin and Wesson \cite{OW} noticed that an observer located at w = const. slice will measure a nonzero stress tensor arising from the 5-dimensional geometry and observers located at different slices acquire different values of the c.c. Similar ideas are develop in this work . The bulk geometry is chosen to have the special form (2.1) for to facilitate a connection between acceleration and $w$. We have used the Shiromizu et al. \cite{SMS} recipe to find the brane-world stress tensor which satisfies the equation of state $\rho = -p = const.<0$ and, therefore, the metric on the brane is of AdS type, with $\Lambda = -3a^{2}$, where $a = 1/w_{0}$. The mathematical form of the proper acceleration led us to read an uniformly accelerating motion as an immersion in the 5th dimension, at $w_{0} = 1/a$.

\end{document}